# Andrei V. Pakoulev[*], Vladimir Burtman[†] and Dmitry Zaslavsky[‡§]

# Time-Resolved Measurements of the Interfacial Charge Transfers by Capacitive Voltage Probe.

**Abbreviations:**
1,8-naphthalene-dicarboxylic anhydride-4, 5- dicarboxylic imide (see also figure 1)
1,4,5,8-naphthalene-tetracarbpxylic diimide

**Manuscript Information:**
9 text pages (including the title page, legends and references)
4 figures

[*] University of Wisconsin-Madison, Department of Chemistry
[†] University of Utah, Department of Physics, Department of Geophysics
[‡] Independent Scientist
[§] To whom the correspondence should be addressed: email: dzaslav@gmail.com P.O. Box 552, Upton, NY, 11973 Phone: (217) 390-3684

SUMMARY (147 words)

**When redox active molecules are covalently tethered to the surface of the n-doped silicon, light induces their oxidation by the semiconductor. To visualize this charge separation we have followed the formation and decay of the surface photovoltage by an auxiliary transparent electrode, which serves as a capacitive probe that picks up the transient surface photovoltage. This method has been used in electronics and biophysics but is novel for the surface redox chemistry. The probe provides both the direction of the charge transfer and the resolution of the dynamics. Its application is independent of the optical properties of the semiconductor. Laser initiation of the reaction permits quantitative measurements of the reactions with the rate constants $\sim 10^8$-$10^3$ s$^{-1}$. The electron injection from covalently tethered NTCI(A)$^{**}$ molecules into silicon was fast (k $>10^9$ s$^{-1}$). The back electron transfer is poly-disperse with the components ranging from ~10 microseconds and up.**

---

$^{**}$ 1,8-naphthalene-dicarboxylic anhydride-4, 5- dicarboxylic imide (see also figure 1)

INTRODUCTION

Interfacial electron transfer plays a key role in molecular electronics and biophysics. For many such systems the time resolved UV-vis spectroscopy has greatly helped to identify and describe the electron transfer reactions involved in charge separation (surface polarization). The UV-visible spectroscopy has become predominant method used in the studies of various electron transfer systems. In particular it has been extensively utilized in dynamic studies of the dye-sensitized photovoltaic cells [1,2].
In these cells it was shown that the energy of light converts directly into the energy of separated charges via injection of electrons from the excited dye molecules into the conductance band of titanium oxide, which transmits the visible light. The whole sequence of the redox transfers has been revealed by the time-resolved spectroscopy.

It is important to note that light induced charge separation is not the only known mechanism for the energy transduction at the surface. The second well known mechanism has been observed in the natural photosynthesis. In this case the light does not cause immediate charge separation between redox centers. Instead the energy travels to the photosynthetic reaction centers via exciton migration between the light-harvesting antennae molecules.

Recently we have reported a light harvesting mechanism which is apparently somewhat similar to the natural photosynthesis because it involves spatial energy transfer[3]. We have used silicon surface with covalently tethered molecules of NTCDI[††] (naphthalene tetracarboxylic diimide), which formed a dense monolayer. A small silver drop that covered only a very small fraction of the chip has been used as a counter-electrode. In this system the energy of light converts into electricity despite the light illuminates the surface several millimeters away from the edge of the silver electrode.

Previously we have posited that the conservation and spatial transfer of energy along the surface is achieved via the charge separation across the junction (**Figures 1A and 1B**) followed by migration of these induced dipoles (see Ref. 3 for details). This mechanism is different from the migration of exitons involved into the natural photosynthesis. Nevertheless, this exciton migration, even though unlikely, has not been experimentally ruled out. While the ruthenium dyes adsorbed at the surface of transparent TiO2 are convenient for the spectroscopic studies, the silicon/NTCDI system is not due to the strong absorption of silicon in the visible region.

In this study we present a novel method for time-resolved studies of interfacial redox reactions. It is suitable for various interfaces irrespectively of their optical properties. Here we used the surface photovoltage technique, which is sensitive to the net charge displacement [4,5], the most basic property of redox reactions. The approach is based on the following: since the spatially organized charge transfers displace charges perpendicular to the junction (**Figure 1A**), this transient displacement can be picked up by a capacitive probe.
The original idea of capacitive probe sensitive to the induced changes in polarization of the medium between the capacitor planes has been proposed by Bergmann[6] in 1932. The method is very convenient when the transient polarization is induced by light coming through a transparent auxiliary electrode. In the semiconductor technology the capacitive probe has been used to study the photovoltage generated in p-n junctions for the quality control [7,8]. This nondestructive non-contact method is able to measure the lifetimes of the minority charge carriers [9,10]. Such relatively slow processes (up to $10^{-5}$ s) can be studied using a chopped light beam. Much faster reactions have been studied by capacitive probe in chloroplasts where nanosecond kinetics of primary charge separations have been resolved using pulsed lasers instead of the chopped beam [11,12].

---

[††] 1,4,5,8-naphthalene-tetracarbpxylic diimide

To the best of our knowledge the capacitive probe technique has never been applied to surface redox chemistry. We show that the capacitive probe measurements provide both the direction of the charge transfer and the information about the charge separation and the kinetic of its recombination.

MATERIALS AND METHODS

The silicon chips functionalized with amino groups were prepared and left as is or topped with the either sparse or dense layers of NTCI(A) as described previously [3,13]. The electrometric measurements were performed with a 200 nm thick ITO electrode deposited on a glass plate and separated from the NTCDI film on the silicon surface by a 0.026 mm insulating spacer (**figure 1B**). The capacities of the sandwiched samples were measured with a Keithly voltmeter. The input resistor of the oscilloscope was 1 MOhm. In the experiments with an external 1:10 voltage splitter the apparent input resistance was 10 MOhm. In these experiments a 1/10 of the actual voltage has been applied to the input channel. The output characteristics of the Ti/sapphire laser were: 800 nm, 1 ps duration and 10 μJ per flash power, 1 kHz or 200 Hz repetition rate. Note that NTCI(A) does not absorb at 800 nm in contrast to 532 nm utilized in the previous work. The signals were recorded by an averaging Tectronics oscilloscope (512 traces) recording up to 2 GS/second.

Figure 1A illustrates the putative electron transfer reactions measured by the capacitive probe upon illumination of our samples. The energy levels of silicon and NTCDI molecules are aligned accordingly to the previously established fact that the light is absorbed by silicon (Figure 1A, the left panel). After an electron crosses from the valence band into the conductance band, this hole is refilled by extraction of an electron from an NTCDI molecule. As a result of the charge separation between silicon and the organic layer a surface dipole is formed (right panel). The life-time of this dipole is finite, and its formation (rate $k_s$) is followed by its slower back-recombination (rate $k_{br}$). The right panel shows the proposed structure of the cation-radical formed upon oxidation of the neutral organic molecule.

BACKGROUND OF THE METHOD

The method measures the transient voltage drop between the plates of the capacitor formed by the semiconductor chip and a probe ITO electrode transparent to the visible light. Under even illumination the density of the charges is the same throughout the film plane. Then the film becomes a surface of equal potential and can be considered as a plate of a complex 3-plate capacitor. Its planes are depicted in the **figure 1B**: the surface of the silicon chip (plane 1), the monolayer of the redox active molecules (plane 2) and the ITO surface (plane 3) separated from plane 2 by a 25 micrometer insulating spacer. In the dark the charge separation between the silicon and ITO planes results from the difference in the work functions. However, this constant field causes no current through the resistor and only the field from the light-induced charges must be considered. When cation-radicals form, the electrostatic field between the planes 1 and 2 changes and a voltage appears between the plates 1 and 3. This voltage creates an electric current through the outer resistance (oscilloscope) until it is compensated by the charge exchange between planes 1 and 3. At any given time the charge conservation law gives:

$$q_1 + q_2 + q_3 = 0 \tag{1}$$

Denoting the area of the plates as S, disregarding the edge effects and utilizing the expression for the electric field of a plane and the field superposition principle we obtain the following for the voltage drop between the plates 1 and 3:

$$V_{1,3} = \varphi_3 - \varphi_1 = +\frac{q_1 d_1}{2\varepsilon_1 \varepsilon_0 S} - \frac{q_2 d_1}{2\varepsilon_1 \varepsilon_0 S} - \frac{q_3 d_1}{2\varepsilon_1 \varepsilon_0 S} + \frac{q_1 d_3}{2\varepsilon_3 \varepsilon_0 S} + \frac{q_2 d_3}{2\varepsilon_3 \varepsilon_0 S} - \frac{q_3 d_3}{2\varepsilon_3 \varepsilon_0 S} \tag{2}$$

Since $q_1 = -q_2 - q_3$, this field represents the superposition of the two fields within the capacitive structures: first capacitor formed by the planes 1 and 2 with the charge $q_2$, second, formed by the planes 1 and 3 with the charge $q_3$ and two layers of dielectric:

$$V_{1,3} = -q_2 \frac{d_1/\varepsilon_1}{\varepsilon_0 S} - q_3 \frac{d_1/\varepsilon_1 + d_3/\varepsilon_3}{\varepsilon_0 S} \tag{3}$$

Denoting $\frac{d_1/\varepsilon_1}{\varepsilon_0 S} = \frac{1}{C_{12}}$, $\frac{d_1/\varepsilon_1 + d_3/\varepsilon_3}{\varepsilon_0 S} = \frac{1}{C_{13}}$ and applying the Ohm's law we obtain:

$$\frac{\partial q_3}{\partial t} + \frac{q_3}{RC_{13}} = -\frac{q_2}{RC_{12}} \tag{4}$$

The negative sign before the electric current indicates that the plane 3 charges oppositely to the plane 2. The general solution of this inhomogeneous equation is a sum of the general solution of the corresponding homogeneous equation and a particular solution of the inhomogeneous equation. Let us consider an important case of the "instantaneous" charge separation and its monodisperse recombination. Then the inhomogeneity is exponential: $q_2 = q_0 e^{-kt}$. A straight forward substitution shows that:

$$q_3 = \frac{q_0 C_{13}/C_{12}}{kRC_{13} - 1} e^{-kt} \tag{5},$$

satisfies equation 4, when $1/k \neq RC_{13} = \tau$, the instrument time-constant. Finally, substituting the initial condition $q_3|_{t=0} = 0$ we obtain the following:

$$q_3 = \frac{q_0 C_{13}/C_{12}}{k\tau - 1} (e^{-kt} - e^{-t/\tau}) \tag{6}.$$

We obtain the expression for the signal measured by the oscilloscope from the Ohm's law and equation 6:

$$V_{1,3} = -R\frac{\partial q_3}{\partial t} = \frac{q_0/C_{12}}{k\tau - 1}(k\tau e^{-kt} - e^{-t/\tau}).$$

Denoting $q_0/C_{12} = V_0$ we obtain the final expression for the signal:

$$V_{1,3} = V_0 e^{-kt} + \frac{V_0}{k\tau - 1}\left[e^{-kt} - e^{t/\tau}\right] \tag{7}$$

Expression (7) shows that a monodisperse charge recombination gives rise to a bi-exponential signal. Let us consider rapid recombination of the dipoles: $k\tau \gg 1$.

The first additive term of equation (7) reflects the kinetics of the surface charge displacements. The second term is the deviation of the measured signal from the charge dynamics. It is important that the transient crosses the V=0 level before recovering the base line. Such behavior is an intrinsic property of the RC chain and reflects the simple fact that the probe **charges and then discharges** during the measurements, while the oscilloscope measures the **current** through its input resistance. Hence, the second component reflects the input resistance of the oscilloscope and the capacity of the sample/probe capacitor and therefore depends on the size of the sample.

When $k \ll RC$, the rapidly formed voltage over the resistor $V_0$ will dissipate as a result of charge redistribution between the plates 1 and 3. This component will reflect the instrumental time-constant $\tau = RC$. In our experiments it was in the range of 0.2-2 ms. Similarly, the rate of the base line recovery will correspond to the time constant of the charge recombination.

In the most general case of heterogeneous surface signal the expression for the $q_2$ becomes the following:

$$q_2(t) = \int_0^\infty \sigma(k) e^{-kt} dk .$$

The appropriate analytical expression for the signal shape can be obtained by appropriate integrating of the equation 7. Remember that in the special case $k\tau = 1$ the deviation term from the equation 7 becomes $DT = kt\, e^{-kt}$.

RESULTS AND DISCUSSION

A typical photovoltage signal from the sparse film of the NTCI(A) molecules is shown in the **figure 2**. It is complex and includes "instantaneous" photovoltage onset, its heterogeneous decay below V=0 and recovery of the base line. The onset of the voltage was not resolved by the oscilloscope and therefore must be faster than 1 ns. Its polarity corresponds to charging the NTCI(A) film positively. This observation perfectly agrees with our previous assumption that silicon photo-oxidizes the neutral NTCI(A) molecules, which form cation-radicals upon ejection of electrons into the semiconductor [3].

Minor electric responses are observed even in the absence of the NTCI(A) coating. Bare silicon surfaces are often contaminated with electrochemically active impurities. Such "as is" surfaces give highly irreproducible electric responses, which strongly decrease after rinsing the surface with acetonitrile (the data are not shown). This artifact is almost eliminated after $NH_2$-functionalization of the silicon surface (**figure 2**). The remaining signal could originate from certain polarization of the silicon wafer due to introduction of the gradient of the minority charge carriers near the surface [14]. Another possibility is the light-induced change of the built-in potential [15]. Regardless of its nature, this artifact is minor; it has an **opposite** polarity and therefore cannot account for the main signal. No signal is observed when the surface of silicon is protected by a thick (>200 nm) insulating layer of $SiO_2$. The insulating oxide layer prevents both the interfacial redox chemistry and band bending.

While the initial charge separation is too fast to be resolved, the heterogeneous recombination is clearly resolved. The fast phase is independent of the input resistance and has a time constant of ~ 10 microseconds. Its contribution to the signal decay varies among the samples and is about 1/3 of the total amplitude of the signal presented in **figure 2**. The slow phase demonstrates a clear proportionality to the input resistance and stretches 10-fold upon addition of a 9 MOhm additional resistor. Therefore the baseline recovery likely represents re-reduction of the slowest cation-radicals. This recovery component is difficult to measure because its amplitude is small.

It is important to note that in the dense NTCI(A) films the contribution of the slow component of the photovoltage decay does not exceed 10% (**figure 3**). This observation agrees well with our previous conclusion that in the dense films the molecules rapidly exchange the oxidizing equivalent with their neighbors [3]. Therefore, the positive charges formed in the "slow" molecules drain back to the silicon substrate via the two-step process: (i) exchange with the "rapid" molecules within interconnected redox network and (ii) the following discharge of these "rapid" molecules. Overall this process is apparently faster than the direct discharge of the "slow" molecules. It is also important to note that these "slow" NTCI(A) molecules can be pre-charged in the discontinuous networks with the intense constant light (**figure 4**). A strong continuous illumination decreases the total amplitude of the signal, mainly at the expense of the slow component while the fast component remains essentially unchanged. Despite lacking the appropriate equipment to quantify the effects of the constant light, the qualitative results are clear. The heterogeneity of the recombination probably reflects the micro-heterogeneity introduced by the silicon cleaning procedures and/or the microscopically uneven thickness of the siloxane layer and is beyond the scope of the present work. Similar kinetic heterogeneity has been observed in PTCDI-doped SAMs grown on ITO by single molecule fluorescence analysis [16] and has been attributed to the differences in the local environment.

What is important to us is that in contrast to the photosynthetic light-harvesting antennas, the energy of light is conserved by the black cells via interfacial electron transfer. The silicon/film junction forms an electrochemical capacitor charged by the light. This capacitor stores both the electrostatic energy and the free energy of the back recombination [17]. It is important to be able to measure and control the discharge characteristics of this capacitor in order to utilize it as an energy source. The electric field formed within this capacitor is measured directly by the auxiliary electrode.

The time resolution limits of capacitive probing are defined by instrumental limitations. A typical 2G-sample/second oscilloscope conveniently measures changes of ~ $10^8 s^{-1}$ (20 points per phase). The reactions that are slower than the instrumental time-constant RC are also more difficult to measure. Therefore the ideal dynamic range for the capacitive measurements is between $10^8$-$10^3$ $s^{-1}$.

Capacitive probing has a lower time-resolution than the pump-probe spectroscopy[18] but it does avoid some of the pump-probe limitations. First, capacitive probing is independent of the optical properties of the semiconductor. Second, it works with flat surfaces which contain fewer molecules than nano-particle suspensions [18,19]. Indeed, separation of only $5.5 \times 10^9$ charges per $cm^2$ across ~ 20 Å will induce a strong signal with amplitude of ~ 1 mV (for $\varepsilon \approx 2$). This number of molecules dissolved in 1 $cm^3$ corresponds to <10 pM concentration, far below the sensitivity limit of any spectroscopic method except for fluorescent techniques [16]. However, the electrochemical measurements work even for non fluorescent molecules and they provide nanosecond time-resolution. The technique also provides the direction of the charge transfer.

**Conclusion**

We develop a relatively simple non-spectroscopic approach to probe the redox dynamics on the surface of semiconductor coated by organic films. The method appears to be much more sensitive than spectroscopic techniques widely used in the surface chemistry. Understanding of the charge transfer through interfaces and its dynamics is crucial for the design of electronic and optoelectronic devices based on organic molecular films and, especially, monomolecular layers. Light controlled capacitive probe is a non-contact non-destructive method and may be applied to study the wide range of such devices, including organic light-emitted diodes, organic solar sells and characterization of organic thin films.

**Acknowledgements**: This work was supported in part by the National Science Foundation under grant CHE-0650431. Acknowledgement is also made to the donors of the Petroleum Research Fund of the American Chemical Society for their support of this research. We thank Dr. John Wright for helpful comments and his kind permission to use the laser equipment and his comments on the manuscript.

**Figure Legends:**

**Figure 1.**
**a)** Light induced charge transfer at the silicon-NTCI(A) interface. The relative arrangement of the energy levels of silicon and NTCI(A) is necessitated by the observed polarity of the photo-induced charge transfer and by the fact that the light is absorbed by silicon. The structure of the putative cation radical is presented on the left. Vector D stands for the induced electric displacement.
**b)** The capacitive probe measurements. The silicon substrate forms the capacitor plane 1. The neutral NTCDI(A) molecules and the putative cation-radical formed upon illumination form the imaginative plane 2. This organic layer is separated from the tin-doped indium oxide electrode (plane 3) by a 25 micrometer insulating film. The capacitor planes 1 and 3 are connected respectively to the ground and signal input of the oscilloscope (V), which has a standard input resistance of 1 MOhm.

**Figure 2.** Light-induced voltage transients. Samples were used as capacitor faces separated from the ITO probe electrode by a 25 micrometer insulating spacer. The voltage transients represent polarization and depolarization of the Si/film junctions: Si/NTCDI(A) (solid line), Si/SiO2 (dotted line), Si surface derivatized with amino groups (dashed line)

**Figure 3.** Charge recombination in connective and sparse films. Dense and sparse films were prepared as described in materials and methods. The samples with dense film of NTCI(A) (dashed line) and the sparse film (solid line) were included as capacitor faces as described in the text and figure 1.

**Figure 4.** Influence of the constant light. Photovoltage transients were without (solid line) and with background continuous illumination by a 100 W bulb placed 40 cm away from the sample (dashed line).

**Figure 1**

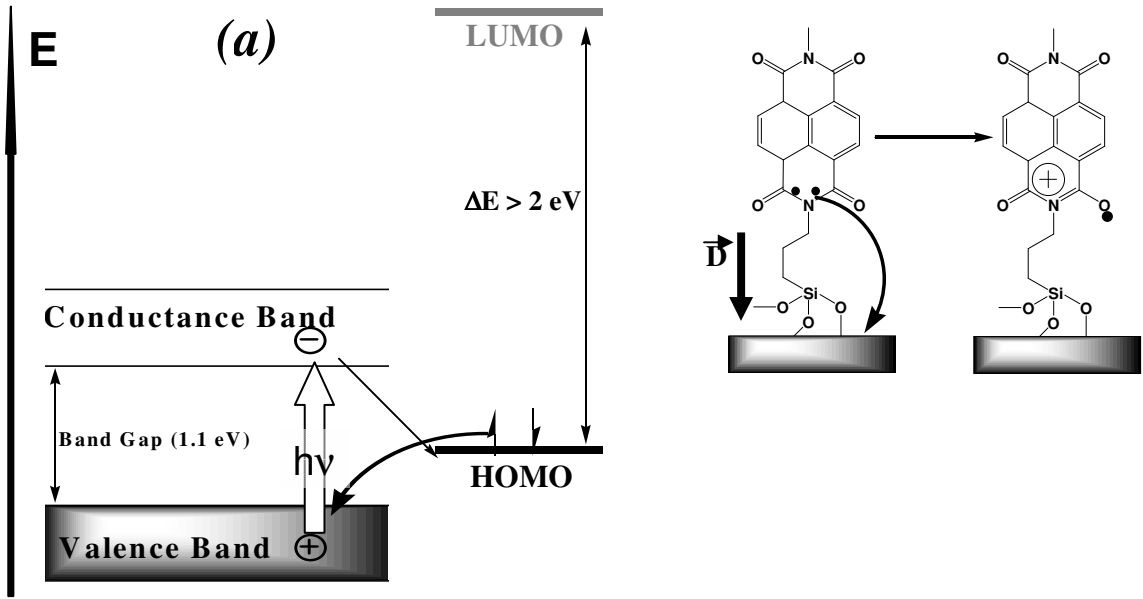

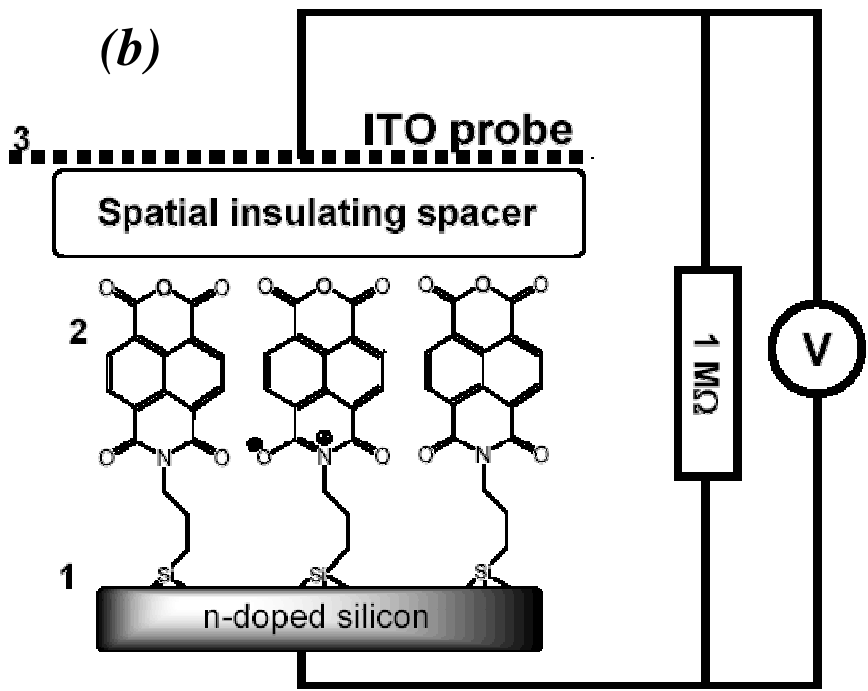

Figure 2.

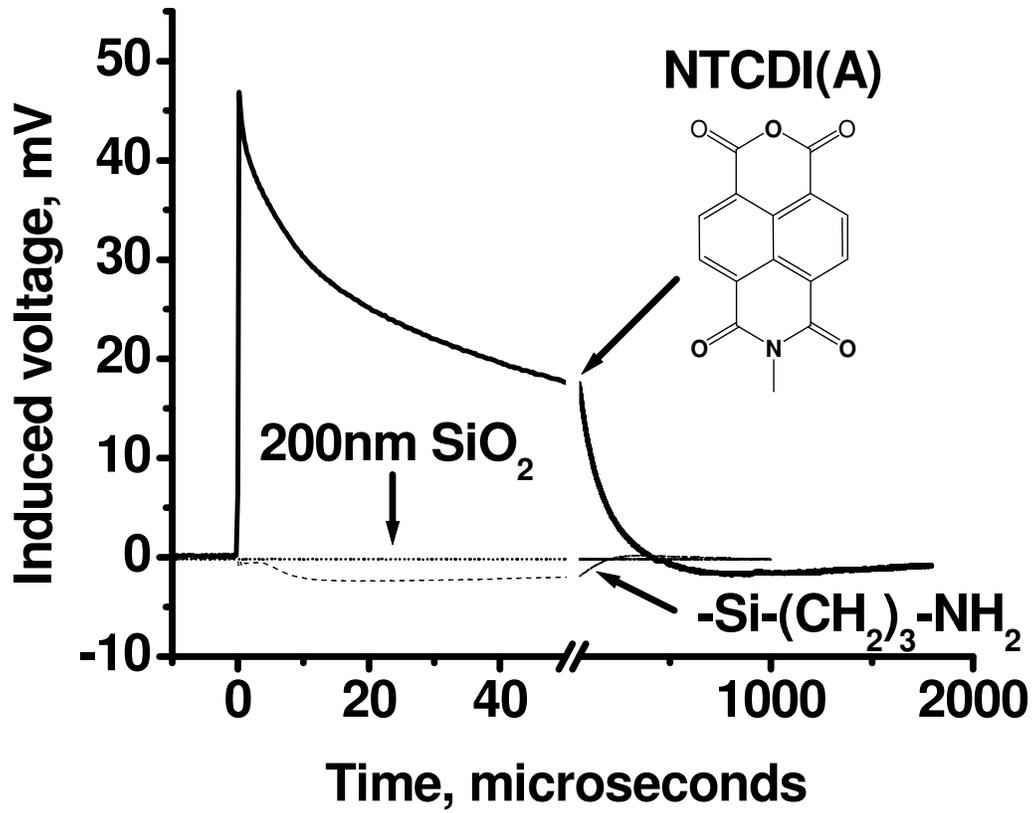

**Figure 3**

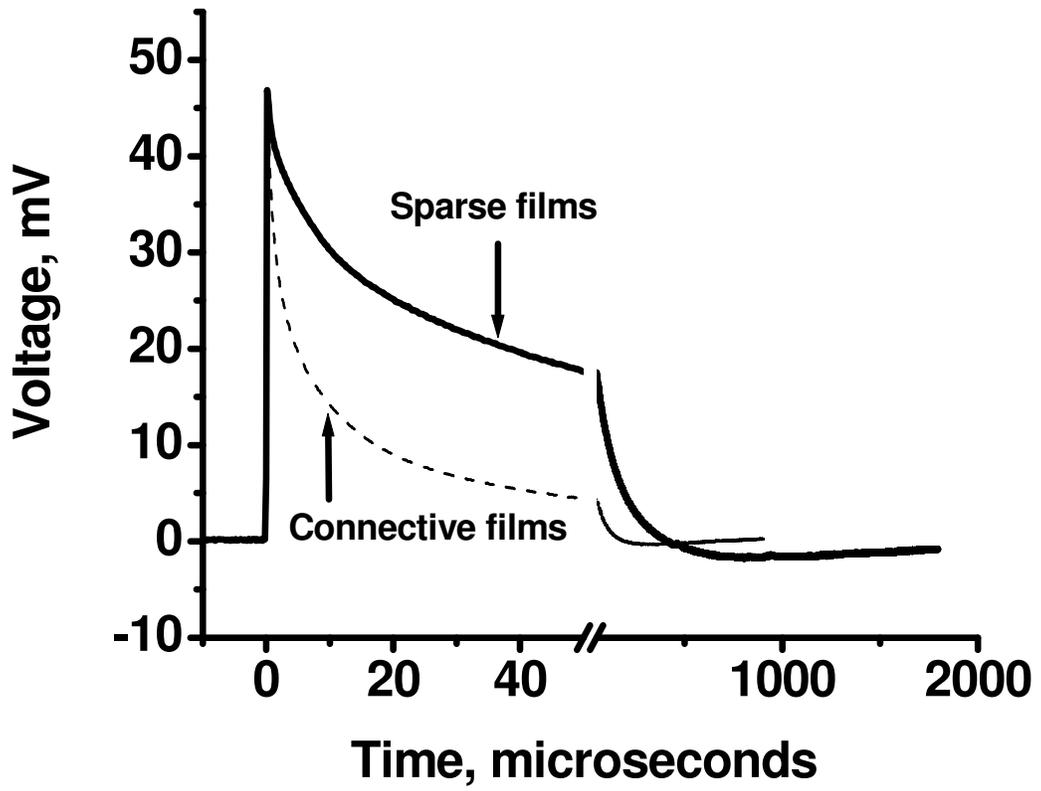

**Figure 4**

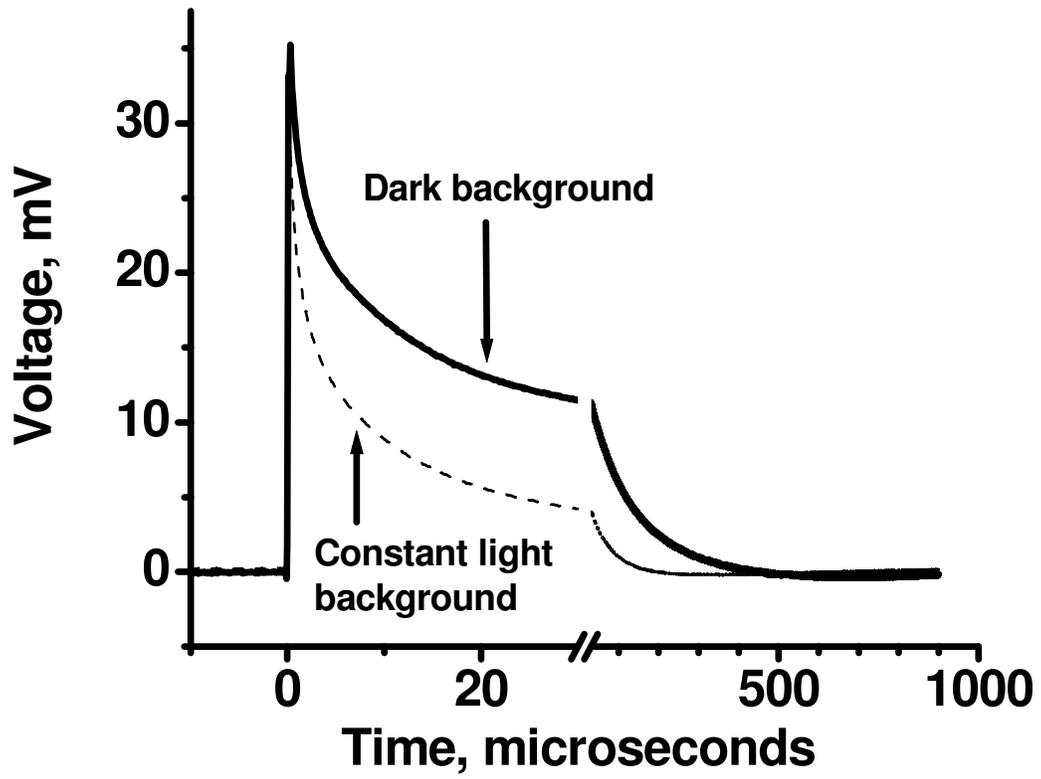